\documentclass[english,aps,reprint,twocolumn,superscriptaddress]{revtex4-1}
\usepackage[T1]{fontenc}
\usepackage[latin9]{inputenc}
\usepackage{ulem}
\usepackage{amstext}
\usepackage{amssymb}
\usepackage{esint}
\usepackage{graphicx}
\newcommand{\eqref}[1]{(\ref{#1})}
\usepackage{babel}
\usepackage{color}

\begin{document}
\title{Emission spectral non-Markovianity in qubit-cavity systems in the ultrastrong coupling regime}
\author{Chenyi Zhang}
\affiliation{Department of Physics, School of Science, Zhejiang University of Science and Technology, Hangzhou 310023, China}
\author{Minghong Yu}
\affiliation{Department of Physics, School of Science, Zhejiang University of Science and Technology, Hangzhou 310023, China}
\author{Yiying Yan}\email{yiyingyan@zust.edu.cn}
\affiliation{Department of Physics, School of Science, Zhejiang University of Science and Technology, Hangzhou 310023, China}
\author{Lipeng Chen}
\affiliation{Max Planck Institute for the Physics of Complex Systems, N\"{o}thnitzer Str., 38, 01187 Dresden, Germany}
\author{Zhiguo L\"{u}}
\affiliation{Key Laboratory of Artificial Structures and Quantum Control
(Ministry of Education), School of Physics and Astronomy,
Shanghai Jiao Tong University, Shanghai 200240, China}
\author{Yang Zhao}\email{YZhao@ntu.edu.sg}
\affiliation{Division of Materials Science, Nanyang Technological University, Singapore 639798, Singapore}
\begin{abstract}
We study the emission spectra of dissipative Rabi and Jaynes-Cummings models in the non-Markovian and ultrastrong coupling regimes. We have derived a polaron-transformed Nakajima-Zwanzig master equation (PTNZE) to calculate the emission spectra, which eliminates the well known limitations of the Markovian approximation and the standard second-order perturbation. Using the time-dependent variational approach as benchmark, the PTNZE is found to yield accurate emission spectra in certain ultrastrong coupling regimes where the standard second-order Nakajima-Zwanzig master equation breaks down. It is shown that the emission spectra of the dissipative Rabi and Jaynes-Cummings models are in general asymmetric under various initial conditions. Direct comparisons of spectra for the two models illustrate the essential role of the qubit-cavity counter-rotating term and the spectra features under different qubit-cavity coupling strengths and system initial conditions.
\end{abstract}
\date{\today}
\maketitle

\section{introduction}
Ultrastrong light-matter coupling has been realized in the context of artificial atoms such as
superconducting qubits, semiconductor quantum wells, and hybrid quantum systems \cite{Forn_D_az_2019,Frisk_Kockum_2019,Sun1}.
The ultrastrong coupling regime is reachable between artificial atoms and discrete modes of electromagnetic field in a resonator~\cite{Forn_D_az_2010,Yoshihara_2017,Wang_2020} as well as between artificial atoms and the electromagnetic continuum in a waveguide~\cite{Forn_D_az_2016}.
In the former case, the light-matter interaction strength is comparable with the bare frequencies of the system, while in the latter case, the emission rate of the artificial atom is in the same order of magnitude as the atomic bare transition frequency. The ultrastrong light-matter coupling leads to the formation of highly hybridized light-matter state, e.g., polaritons, which become increasingly relevant in novel chemistry and quantum technologies such as quantum information processing~\cite{Blais_2021}, photochemistry~\cite{Ulusoy_2019,Sun2,Sun3}, control of chemical reaction rates~\cite{Herrera_2016,Mart_nez_Mart_nez_2017,Sidler_2020,Saller_2021,Lindoy_2022}, and polariton mediation of electron transfer~\cite{Mandal_2020}.

There are several consequence of the ultrastrong coupling between the artificial atom and the electromagnetic continuum.
First, the Lamb shift, or the renormalization of the bare transition frequency of the artificial atoms can be considerable~\cite{Forn_D_az_2016}.
Second, the counter-rotating terms cannot be neglected, which leads to the breakdown of the rotating-wave approximation (RWA), an entangled ground state dressed by virtual photons of the coupled system,
and the nonconservation of the excitation number of the entire system~\cite{Cirio_2016,De_Liberato_2017}.
Third, the time evolution of the atoms is generally no longer Markovian.
This results in the failure of the widely used Markovian master equation as well as the Markovian quantum regression theory.
In addition, the standard second-order perturbation used in the master equation is also rendered questionable in the regime of ultrastrong coupling.
It is then challenging to calculate the emission spectrum of the artificial atoms due to the inapplicability of the widely used approximations in the ultrastrong coupling regime.

Capable to describe a qubit ultrastrongly coupled to a single cavity mode and a radiation reservoir, a scenario easily realizable in superconducting circuits, the dissipative quantum Rabi model provides a unique platform for testing various phenomena associated with the ultrastrong coupling.
This model has been treated by several methods in the ultrastrong coupling regime such as the path integral approach~\cite{Henriet_2014},
the method of the dynamical polaron ansatz and the matrix product state~\cite{Zueco_2019}, and the time-dependent variational approach with the multiple Davydov ansatz~\cite{Yan_2021}. In particular, the non-Markovian dynamics and the spontaneous emission spectrum of the model has been carefully examined in the regime of ultrastrong coupling.
However, most authors are concerned with the single-excitation case, i.e., the issue of vacuum Rabi splitting.
Few efforts have been devoted to the multiple-excitation cases.
In the presence of intra-cavity photons, for example, emission spectra of a qubit-cavity system have not been studied in the regime of ultrastrong coupling.
Furthermore, it remains unanswered how the counter-rotating qubit-cavity coupling alters the emission spectra in the presence of strong dissipation.

In this work, we will examine the emission spectra of the quantum Rabi model and the Jaynes-Cummings (JC) model in a dissipative bath and in the ultrastrong coupling regime by using the non-Markovian quantum regression theory based on
a polaron-transformed Nakajima-Zwanzig master equation (PTNZE) and the variational approach with the multiple Davydov ansatz~\cite{Wang_2016,Wang_2017,Fujihashi_2017,Yan_2021}.
The advantage and the validity of the PTNZE is analyzed by comparing its predictions with those of the variational approach and the standard second-order Nakajima-Zwanzig master equation
(SNZE). It is found that the PTNZE provides a satisfactorily accurate description in certain regimes of ultrastrong coupling where
SNZE fails. The emission spectrum
is found to be asymmetric regardless of the counter-rotating interactions between the qubit and the cavity. Interestingly, for some initial states, the emission spectra from the two models in a dissipative bath show little difference.

The remainder of the paper is organized as follows.
In Sec.~\ref{sec:MM}, we introduce the theoretical frameworks employed
in this work, including the system Hamiltonians
and the methodologies. Results and the corresponding
discussion are given in detail in Sec.~\ref{sec:RD}. Finally, we conclude
in Sec.~\ref{sec:Con}.

\section{Models and methodologies}\label{sec:MM}

We consider the dissipative Rabi model in which a qubit is coupled to a
high-Q cavity mode and a radiation reservoir, which is described by
($\hbar=1$),
\begin{equation}
H=H_{{\rm Rabi}}+H_{{\rm R}}+H_{{\rm I}},\label{eq:htot}
\end{equation}

\begin{equation}
H_{{\rm Rabi}}=\frac{1}{2}\omega_{0}\sigma_{z}+\omega_{c}\hat{b}_{c}^{\dagger}\hat{b}_{c}+\frac{\lambda_{c}}{2}(\hat{b}_{c}^{\dagger}+\hat{b}_{c})\sigma_{x},
\end{equation}
\begin{equation}
H_{{\rm R}}=\sum_{k}\omega_{k}\hat{b}_{k}^{\dagger}\hat{b}_{k},
\end{equation}
\begin{equation}
H_{{\rm I}}=\sum_{k}\frac{\lambda_{k}}{2}(\hat{b}_{k}^{\dagger}+\hat{b}_{k})\sigma_{x}.
\end{equation}
As described by $H_{{\rm Rabi}}$, a qubit of the transition
frequency $\omega_{0}$ interacts with the single cavity mode of frequency
$\omega_{c}$, where $\sigma_{i}$ $(i=x,y,z)$ is the Pauli matrix, $\lambda_{c}$
is the qubit-cavity coupling strength, and $\hat{b}_{c}^{\dagger}$ and
$\hat{b}_{c}$ are the creation and annihilation operator of the cavity,
respectively. $H_{{\rm R}}$ is the free Hamiltonian of the radiation
reservoir, where $\omega_{k}$ is the frequency, $\hat{b}_{k}^{\dagger}$
and $\hat{b}_{k}$ are the creation and annihilation operator, respectively.
$H_{{\rm I}}$ is the interaction between the qubit and reservoir.
$\lambda_{k}$ is the coupling constant between the $k$th mode and
qubit. The dissipative bath is assumed to be of the Ohmic
type, i.e., the bath spectral density is given by
\begin{equation}
J(\omega)=\sum_{k}\lambda_{k}^{2}\delta(\omega_{k}-\omega)=2\alpha\omega\exp(-\omega/\omega_{{\rm cut}}),
\end{equation}
where $\alpha$ is a dimensionless coupling strength and $\omega_{{\rm cut}}$
is the cut-off frequency. In our model, the dissipation of the cavity is assumed to be negligible.

To examine the role of the counter-rotating coupling on the emission spectrum, we also consider the dissipative JC model, which
can be obtained by replacing the Rabi Hamiltonian with the JC Hamiltonian in Eq.~\eqref{eq:htot},
\begin{equation}
 H_{{\rm JC}}=\frac{1}{2}\omega_{0}\sigma_{z}+\omega_{c}\hat{b}_{c}^{\dagger}\hat{b}_{c}+\frac{\lambda_{c}}{2}(\hat{b}_{c}\sigma_{+}+\hat{b}_{c}^{\dagger}\sigma_{-}),
\end{equation}
where $\sigma_\pm=(\sigma_x\pm i\sigma_y)/2$.

We are interested in the spontaneous emission spectrum in the presence of
an initially vacuum radiation bath with zero temperature, which is
defined as the photon number emitted at time $t$ as a function of
the photon frequency, i.e.,
\begin{eqnarray}\label{eq:Nwk}
N(\omega_{k},t) & = & {\rm Tr}[\hat{b}_{k}^{\dagger}(t)\hat{b}_{k}(t)\rho(0)]\nonumber \\
 & = & \frac{\lambda_{k}^{2}}{4}\int_{0}^{t}\int_{0}^{t}\langle\sigma_{x}(t_{1})\sigma_{x}(t_{2})\rangle e^{-i\omega_{k}(t_{1}-t_{2})}dt_{1}dt_{2}.\nonumber \\
\label{eq:sxsx}
\end{eqnarray}
Here the average is taken over the initial density matrix $\rho(0)\equiv\rho_{S}(0)|\left\{ 0_{k}\right\} \rangle\langle\left\{ 0_{k}\right\} |$
with $\rho_{S}(0)$ being the state of the qubit-cavity system and $|\left\{ 0_{k}\right\} \rangle$ being the vacuum state of the reservoir.
The second line of Eq.~\eqref{eq:Nwk} is derived with the Heisenberg representation for $\hat{b}_k(t)$ governed by Hamiltonian \eqref{eq:htot}. Although the defined spectrum is time-dependent, it becomes stable in the long-time limit since the total system is time-independent, namely, the steady-state spectrum can be given by
\begin{equation}
  N(\omega_k)=\lim_{t\rightarrow\infty}N(\omega_k,t).
\end{equation}
Equation~(\ref{eq:sxsx}) indicates two ways to calculate the spectrum. One is the direct evaluation of the photon number of the radiation modes (corresponding to the first line of Eq.~(\ref{eq:sxsx})). The other is based on the Fourier transform of the two-time correlation function (corresponding to the second line of Eq.~(\ref{eq:sxsx})).We first introduce how to calculate the two-time correlation functions by using the nonMarkovian master equation approaches.

The two-time correlation function in the integral can be evaluated as
\begin{eqnarray}
\langle\sigma_{x}(t)\sigma_{x}(t^{\prime})\rangle & = & {\rm Tr}_{{\rm S}}\{\sigma_{x}{\rm Tr}_{{\rm R}}[U(t,t^{\prime})\sigma_{x}\rho(t^{\prime})U^{\dagger}(t^{\prime},t)]\}\nonumber\\
 & \equiv & {\rm Tr}_{{\rm S}}[\sigma_{x}\Lambda_{{\rm S}}(t,t^{\prime})],
\end{eqnarray}
where $U(t,t^{\prime})=\exp[-iH(t-t^{\prime})]$, $\rho(t^{\prime})=U(t^{\prime})\rho(0)U^{\dagger}(t^{\prime})$
is the total density matrix, and $\Lambda_{{\rm S}}(t,t^{\prime})={\rm Tr}_{{\rm R}}[U(t,t^{\prime})\sigma_{x}\rho(t^{\prime})U^{\dagger}(t,t^{\prime})]$
is a reduced operator. Our aim next is to arrive at a time-nonlocal master equation for the reduced operator.

\subsection{Polaron-transformed Nakajima-Zwanzig master equation}
To go beyond the standard second-order perturbation,
an effective Hamiltonian can be derived based on a unitary transformation
$H^{\prime}=e^{S}He^{-S}$ with~\cite{Zheng_2004}
\begin{equation}
S=\sum_{k}\frac{\lambda_{k}}{2\omega_{k}}\xi_{k}(\hat{b}_{k}^{\dagger}-\hat{b}_{k})\sigma_{x},
\end{equation}
where $\xi_{k}$ are undetermined parameters. When $\xi_k=1$, the unitary transformation becomes the standard polaron transformation~\cite{Xu2016}. Omitting the second and higher-order terms in $\lambda_k$ in the transformed Hamiltonian and determining $\xi_k$ by requiring the interaction Hamiltonian to be of a RWA-like form, we arrive at the following effective Hamiltonian,
\begin{equation}
H^{\prime}  \approx  H^\prime_{\rm Rabi}+\sum_{k}\omega_{k}\hat{b}_{k}^{\dagger}\hat{b}_{k}+\sum_{k}\tilde{\lambda}_{k}(\hat{b}_{k}^{\dagger}\sigma_{-}+\hat{b}_{k}\sigma_{+}),\label{eq:Heff}
\end{equation}
where
\begin{equation}
  H^\prime_{\rm Rabi}=\frac{1}{2}\eta\omega_{0}\sigma_{z}+\omega_{c}\hat{b}_{c}^{\dagger}\hat{b}_{c}+\frac{\lambda_{c}}{2}(\hat{b}_{c}^{\dagger}+\hat{b}_{c})\sigma_{x},
\end{equation}
\begin{equation}
\eta=\exp\left[-\sum_{k}\frac{1}{2}\left(\frac{\lambda_{k}}{\omega_{k}}\xi_{k}\right)^{2}\right].
\end{equation}
\begin{equation}
\xi_{k}=\frac{\omega_{k}}{\eta\omega_{0}+\omega_{k}},
\end{equation}
\begin{equation}
\tilde{\lambda}_{k}=\frac{\eta\omega_{0}\lambda_{k}}{\eta\omega_{0}+\omega_{k}}.
\end{equation}
The modification of the coupling constants enables a perturbative treatment in a
moderately strong system-reservoir
coupling. Based on Eq.~\eqref{eq:Heff} and using a standard procedure~\cite{Breuer_2007}, we can obtain PTNZEs of the reduced
operators $\Lambda_{{\rm S}}^{\prime}(t,t^{\prime})={\rm Tr}_{{\rm R}}[e^{S}\Lambda(t,t^{\prime})e^{-S}]$
and $\rho_{{\rm S}}^{\prime}(t)={\rm Tr}_{{\rm R}}[e^{S}\rho(t)e^{-S}]$, which reads
\begin{widetext}
\begin{eqnarray}
\frac{d}{dt}\Lambda_{{\rm S}}^{\prime}(t,t^{\prime}) & = & -i[H_{{\rm Rabi}}^{\prime},\Lambda_{S}^{\prime}(t,t^{\prime})]-\int_{t^{\prime}}^{t}d\tau\left\{ \tilde{C}(t-\tau)[\sigma_{+},e^{-iH_{{\rm Rabi}}^{\prime}(t-\tau)}\sigma_{-}\Lambda_{{\rm S}}^{\prime}(\tau,t^{\prime})e^{iH_{{\rm Rabi}}^{\prime}(t-\tau)}]\right.\nonumber \\
 &  & \left.-\tilde{C}^{\ast}(t-\tau)[\sigma_{-},e^{-iH_{{\rm Rabi}}^{\prime}(t-\tau)}\Lambda_{{\rm S}}^{\prime}(\tau,t^{\prime})\sigma_{+}e^{iH_{{\rm Rabi}}^{\prime}(t-\tau)}]\right\} \nonumber \\
 &  & -\int_{0}^{t^{\prime}}d\tau\left\{ \tilde{C}(t-\tau)[\sigma_{+},\sigma_{x}(t,t^{\prime})e^{-iH_{{\rm Rabi}}^{\prime}(t-\tau)}\sigma_{-}\rho_{{\rm S}}^{\prime}(\tau)e^{iH_{{\rm Rabi}}^{\prime}(t-\tau)}]\right.\nonumber \\
 &  & \left.-\tilde{C}^{\ast}(t-\tau)[\sigma_{-},\sigma_{x}(t,t^{\prime})e^{-iH_{{\rm Rabi}}^{\prime}(t-\tau)}\rho_{{\rm S}}^{\prime}(\tau)\sigma_{+}e^{iH_{{\rm Rabi}}^{\prime}(t-\tau)}]\right\} ,\label{eq:tnlme1}
\end{eqnarray}
\begin{equation}
\frac{d}{dt}\rho_{{\rm S}}^{\prime}(t)=-i[H_{{\rm Rabi}}^{\prime},\rho_{S}^{\prime}(t)]-\int_{0}^{t}d\tau\left\{ \tilde{C}(t-\tau)[\sigma_{+},e^{-iH_{{\rm Rabi}}^{\prime}(t-\tau)}\sigma_{-}\rho_{{\rm S}}^{\prime}(\tau)e^{iH_{{\rm Rabi}}^{\prime}(t-\tau)}]+{\rm h.c.}\right\} ,\label{eq:tnlme2}
\end{equation}
\end{widetext}
where
\begin{equation}
\sigma_{x}(t,t^{\prime})=e^{-iH_{{\rm Rabi}}^{\prime}(t-t^{\prime})}\sigma_{x}e^{iH_{{\rm Rabi}}^{\prime}(t-t^{\prime})},
\end{equation}
\begin{equation}
 \tilde{C}(t)=\int_{0}^{\infty}\left(\frac{\eta\omega_{0}}{\eta\omega_{0}+\omega}\right)^{2}J(\omega)e^{-i\omega t}d\omega.
\end{equation}
Detailed derivation of the PTNZEs can be found in the Appendix~\ref{app:me}.

The PTNZEs are differential-integral equations, which can be solved by introducing a set of auxiliary density matrices~\cite{Meier1999}. To this end, we first fit the bath correlation function by a sum of exponentials~\cite{Duan_2017}. The differential-integral equations are then casted to a set of coupled differential equations, which can be easily integrated by Runge-Kutta algorithm. The detailed numerical treatment of the PTNZE is presented in the Appendix~\ref{app:me}. Finally, the emission spectrum can be obtained by the integral in second line of Eq.~(\ref{eq:sxsx}).
The present treatment is referred to as the non-Markovian quantum regression theory based on the PTNZE.
Following the standard procedure in Refs.~\cite{Goan_2011,McCutcheon_2016}, we can also establish the non-Markovian quantum regression theory based on the polaron-transformed Bloch-Redfield master equation (PTBRE).
The PTBREs are differential equations with time-dependent coefficients, which can be directly integrated with the Runge-Kutta method.
We will address the performance of both PTNZEs and PTBREs.

The present  polaron-transformed master equations  can be applied to the dissipative JC model.
An effective Hamiltonian is first generated by a unitary transformation for the dissipative JC model. Then one derives the PTNZEs for the effective Hamiltonian. In fact, the PTNZEs of the JC model can be obtained from those of the dissipative Rabi model with simple substitutions, and the details of derivation can be found in the Appendix~\ref{app:JCHeff}.

To demonstrate the advantage of the PTNZE, we also consider the SNZE, which is derived based on the original Hamiltonian. In the standard procedure~\cite{Duan_2017}, $H_{\rm Rabi}+H_{\rm R}$ and $H_{\rm I}$ are adopted as the free and the perturbative Hamiltonian, respectively, and the SNZE is derived for
the reduced operators: $\Lambda_{{\rm S}}(t,t^{\prime})$
and $\rho_{{\rm S}}(t)={\rm Tr}_{{\rm R}}[\rho(t)]$, of which the SNZEs are given by
\begin{widetext}
\begin{eqnarray}\label{eq:p2ndme1}
\frac{d}{dt}\Lambda_{{\rm S}}(t,t^{\prime}) & = & -i[H_{{\rm Rabi}},\Lambda_{S}(t,t^{\prime})]-\int_{t^{\prime}}^{t}d\tau\left\{ C(t-\tau)[\sigma_{x},e^{-iH_{{\rm Rabi}}(t-\tau)}\sigma_{x}\Lambda_{{\rm S}}(\tau,t^{\prime})e^{iH_{{\rm Rabi}}(t-\tau)}]\right.\nonumber \\
 &  & \left.-C^{\ast}(t-\tau)[\sigma_{x},e^{-iH_{{\rm Rabi}}(t-\tau)}\Lambda_{{\rm S}}(\tau,t^{\prime})\sigma_{x}e^{iH_{{\rm Rabi}}(t-\tau)}]\right\} \nonumber \\
 &  & -\int_{0}^{t^{\prime}}d\tau\left\{ C(t-\tau)[\sigma_{x},\sigma_{x}(t,t^{\prime})e^{-iH_{{\rm Rabi}}(t-\tau)}\sigma_{x}\rho_{{\rm S}}(\tau)e^{iH_{{\rm Rabi}}(t-\tau)}]\right.\nonumber \\
 &  & \left.-C^{\ast}(t-\tau)[\sigma_{x},\sigma_{x}(t,t^{\prime})e^{-iH_{{\rm Rabi}}(t-\tau)}\rho_{{\rm S}}(\tau)\sigma_{x}e^{iH_{{\rm Rabi}}(t-\tau)}]\right\} ,
\end{eqnarray}
\begin{equation}\label{eq:p2ndme2}
\frac{d}{dt}\rho_{{\rm S}}(t)=-i[H_{{\rm Rabi}},\rho_{S}(t)]-\int_{0}^{t}d\tau\left\{ C(t-\tau)[\sigma_{x},e^{-iH_{{\rm Rabi}}(t-\tau)}\sigma_{x}\rho_{{\rm S}}(\tau)e^{iH_{{\rm Rabi}}(t-\tau)}]+{\rm h.c.}\right\} ,
\end{equation}
\end{widetext}
\begin{equation}
\sigma_{x}(t,t^{\prime})=e^{-iH_{{\rm Rabi}}(t-t^{\prime})}\sigma_{x}e^{iH_{{\rm Rabi}}(t-t^{\prime})},
\end{equation}
\begin{equation}
C(t)=\frac{1}{4}\int_{0}^{\infty}J(\omega)e^{-i\omega t}d\omega=\frac{\alpha\omega_{{\rm cut}}^{2}}{2\left(1+i\omega_{{\rm cut}}t\right)^{2}}.
\end{equation}

\begin{figure}
  \includegraphics[width=\columnwidth]{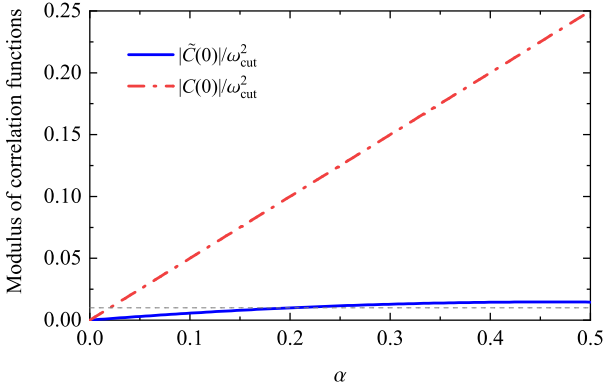}
  \caption{Modulus of correlation functions $\tilde{C}(t)$ and $C(t)$ at time $t=0$ in units of $\omega_{\rm cut}^{2}$ as
  a function of $\alpha$ for $\omega_{\rm cut}=5\omega_0$. The grey dashed line indicates the value $10^{-2}$.}\label{fig1}
\end{figure}

To gain insight into the substantial improvement of PTNZE over SNZE, we analyze the reservoir correlation functions $\tilde{C}(t)$ and $C(t)$
for PTNZE and SNZE, respectively.
Following Ref.~\cite{McCutcheon_2010}, the validity condition of PTNZE can be established as
\begin{equation}
  |\tilde{C}(0)|/\omega_{\rm cut}^2\ll1.
\end{equation}
Similarly, for SNZE, we have the condition
\begin{equation}
  \left|C(0)\right|/\omega_{\rm cut}^2\ll1.
\end{equation}
In Fig.~\ref{fig1}, we show the behaviors of $|\tilde{C}(0)|/\omega_{\rm cut}^2$ and $|C(0)|/\omega_{\rm cut}^2=\alpha/2$ as a function $\alpha$ for $\omega_{\rm cut}=5\omega_0$.
We note that $|\tilde{C}(0)|$ increases extremely slowly with increasing $\alpha$, while $|C(0)|$ increases linearly with increasing $\alpha$.
Roughly speaking, it is expected that the PTNZE is valid for $\alpha<0.2$
while the SNZE can be justified for $\alpha<0.02$.

\subsection{Time-dependent variational principle and multiple Davydov ansatz}
To benchmark the master equation approaches, we use a numerically exact treatment based on the time-dependent variational principle
and the multiple Davydov D$_1$ ansatz to solve the time-dependent Schr\"{o}dinger equation
governed by Hamiltonian \eqref{eq:htot}, i.e., $i\partial_{t}|\psi(t)\rangle=H|\psi(t)\rangle.$
To simplify the calculation, we first transform the Schr\"{o}dinger equation to
the interaction picture associated with the free Hamiltonian of the
reservoir $H_{{\rm R}}$, yielding
\begin{equation}
i\partial_{t}|\tilde{\psi}(t)\rangle=\tilde{H}(t)|\tilde{\psi}(t)\rangle,
\end{equation}
where $|\tilde{\psi}(t)\rangle=\exp(iH_{{\rm R}}t)|\psi(t)\rangle$
and
\begin{equation}
\tilde{H}(t)=H_{{\rm Rabi}}+\sum_{k}\frac{\lambda_{k}}{2}\left(\hat{b}_{k}^{\dagger}e^{i\omega_{k}t}+\hat{b}_{k}e^{-i\omega_{k}t}\right)\sigma_{x}.
\end{equation}
The time-dependent variational principle states that the optimal solution
to the time-dependent Schr\"{o}dinger equation can be found via~\cite{frenkel}
\begin{equation}
\langle\delta\tilde{\psi}(t)|i\partial_{t}-\tilde{H}(t)|\tilde{\psi}(t)\rangle=0\label{eq:tdvp}
\end{equation}
with a given trial state $|\tilde{\psi}(t)\rangle$.

In this work, we use the multiple Davydov D$_1$ (multi-D$_1$) ansatz \cite{wires}, which
takes the form
\begin{equation}
|{\rm D}^{M}_{1}(t)\rangle=\sum_{d=1}^{M}\sum_{i=1}^{N_c}\left[A_{di}|+\rangle|\phi_{i}\rangle|f_{d}\rangle+B_{di}|-\rangle|\phi_{i}\rangle|g_{d}\rangle\right],\label{eq:Ds}
\end{equation}
where $|\pm\rangle$ are the eigenstates of $\sigma_{x}$, $|\phi_{i}\rangle$
are the bases for describing the cavity mode (which can be Fock states, displaced Fock states, and coherent states), and $|f_{d}\rangle$ and $|g_{d}\rangle$ are the multi-mode
coherent states:
\begin{eqnarray}
|f_{d}\rangle & = & \exp\left[\sum_{k=1}^{N_b}\left(f_{dk}\hat{b}_{k}^{\dagger}-f_{dk}^{\ast}\hat{b}_{k}\right)\right]|\left\{ 0_{k}\right\} \rangle,\\
|g_{d}\rangle & = & \exp\left[\sum_{k=1}^{N_b}\left(g_{dk}\hat{b}_{k}^{\dagger}-g_{dk}^{\ast}\hat{b}_{k}\right)\right]|\left\{ 0_{k}\right\} \rangle.
\end{eqnarray}
Here $A_{di}$, $B_{di}$, $f_{dk}$, and $g_{dk}$ are
the time-dependent variational parameters. The physical significance
of these parameters are clear: $A_{di}$ and $B_{di}$ are the probability
amplitudes, and $f_{dk}$ and $g_{dk}$ are the displacements of the $k$th
bath mode.

Formally, there are totally $2MN_{c}$ bases and $2M(N_{b}+N_{c})$ variational parameters in the ansatz.
In this work, we use the Fock bases for the cavity mode. In doing so, we can easily tackle the initial condition that the cavity is initially in a Fock state. Furthermore, the present ansatz can be also applied to the dissipative JC model. One can readily derive the equations of motion for
the variational parameters by substituting Eq.~(\ref{eq:Ds}) into
Eq.~(\ref{eq:tdvp}),
The explicit forms of the equations of motion and the details of the numerical implementation
are presented in the Appendix~\ref{app:eom}.

\begin{figure*}
  \includegraphics[width=2\columnwidth]{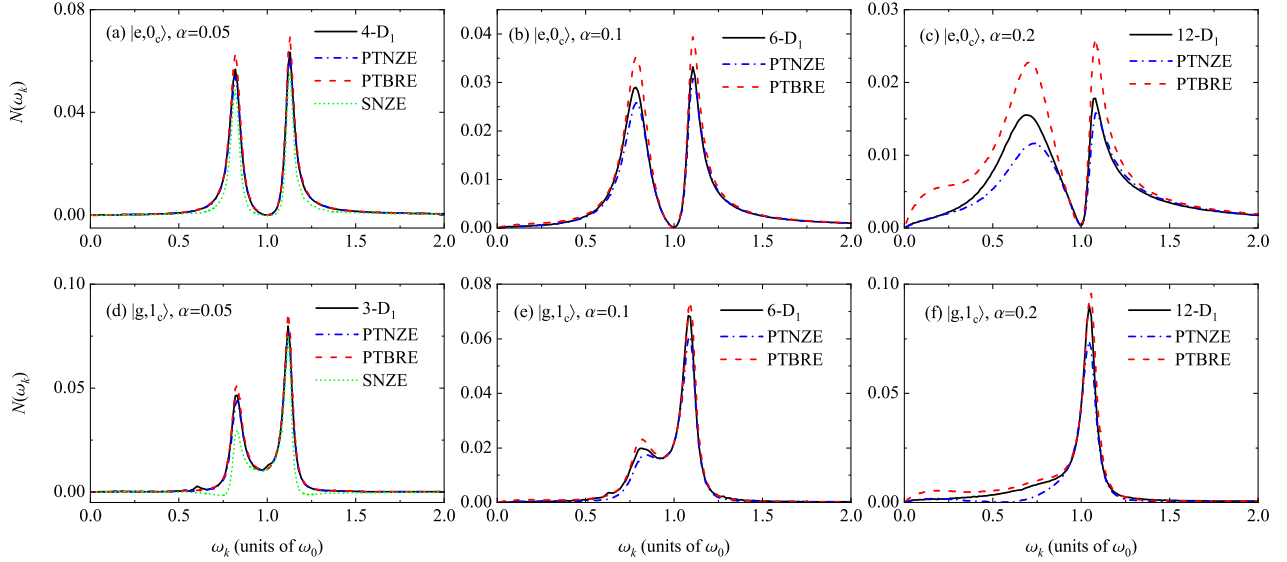}
  \caption{Steady-state emission spectra of the dissipative Rabi model calculated by the variational approach and
  the master equation approaches for $\lambda_c=0.3\omega_0$, $\omega_c=\omega_0$ and the three values of $\alpha$.
  The qubit-cavity initial states $|e,0_c\rangle$ and $|g,1_c\rangle$ are used in the upper and lower panels, respectively.}\label{fig2}
\end{figure*}

In the numerical simulation, we consider a finite number of the harmonic
oscillators of the reservoir. Their frequencies $\omega_{k}$ and
coupling constants $\lambda_{k}$ can be obtained from the discretization
of the spectral density function. We follow the discretization procedure
in Ref.~\cite{Wang_2001} and derive the discretized frequencies in the frequency domain
$(0,\,\omega_{{\rm max}})$,
\begin{equation}
\omega_{k}=-\omega_{{\rm cut}}\ln\left[1-\frac{k}{N_{b}}(1-e^{-\omega_{{\rm max}}/\omega_{{\rm cut}}})\right],
\end{equation}
with the corresponding coupling constants
\begin{equation}
\lambda_{k}=\left[2\alpha\omega_{k}\omega_{{\rm cut}}(1-e^{-\omega_{{\rm max}}/\omega_{{\rm cut}}})/N_{b}\right]^{1/2},
\end{equation}
where $N_{b}$ is the total number of the reservoir modes. Throughout this work, we set $N_b=500$, $\omega_{\rm cut}=5\omega_0$, and $\omega_{\rm max}=4\omega_{\rm cut}$.

With the variational approach, the emission spectrum can be calculated directly from the single-time expectation as follows:
\begin{eqnarray}
N(\omega_{k},t) & = & \langle {\rm D}^{M}_1(t)|\hat{b}_{k}^{\dagger}\hat{b}_{k}|{\rm D}^{M}_1(t)\rangle\nonumber \\
 & = & \sum_{m,d=1}^{M}\sum_{j=1}^{N}\left[A_{mj}^{\ast}A_{dj}f_{mk}^{\ast}f_{dk}\langle f_{m}|f_{d}\rangle\right.\nonumber\\
 &   &\left.+B_{mj}^{\ast}B_{dj}g_{mk}^{\ast}g_{dk}\langle g_{m}|g_{d}\rangle\right],
\end{eqnarray}
where $\langle f_{m}|g_{d}\rangle$ is the overlap between the coherent
states and is given by
\begin{equation}
\langle f_{m}|g_{d}\rangle=\exp\left[\sum_{k=1}^{N_b}\left(f_{mk}^{\ast}g_{dk}-\frac{1}{2}\left|f_{mk}\right|^{2}-\frac{1}{2}\left|g_{dk}\right|^{2}\right)\right].
\end{equation}
The spectra obtained in this manner are referred to as the multi-$D_1$ results.

The accuracy of the variational results can be measured by the scaled
squared norm of the deviation vector~\cite{wires,SLZ_2010,Martinazzo_2020}
\begin{eqnarray}
\sigma^{2}(t) & = & \left|[i\partial_{t}-\tilde{H}(t)]|D^{M}_{1}(t)\rangle\right|{}^{2}/\omega_{0}^{2}\nonumber \\
 & = & \omega_{0}^{-2}\left[\langle D^{M}_{1}(t)|\tilde{H}^{2}(t)|D^{M}_{1}(t)\rangle-\langle\dot{D}^{M}_{1}(t)|\dot{D}^{M}_{1}(t)\rangle\right].\nonumber\\
\end{eqnarray}
Roughly speaking, when $\sigma^{2}(t)<10^{-2}$, the variational results
are numerically accurate~\cite{Yan_2020}. For the present dissipative Rabi and JC models,
the trial state is capable to yield a deviation vector with a tiny magnitude, i.e., $\sigma^2(t)\sim10^{-3}$, with a sufficiently large multiplicity $M$ in the most cases of the single-excitation initial states. Detailed calculation and behavior of  $\sigma^{2}(t)$ as a function of $t$ for various parameters and the two kinds of single-excitation initial states are presented in Appendix~\ref{app:eom}.
Specifically, the present multiple Davydov D$_1$ ansatz can be used in generating the results in Ref.~\cite{Yan_2021}.
Thus the variational approach with the multiple Davydov D$_1$ ansatz is chosen to benchmark the master-equation results in this work.
In the case of the multiple-excitation initial states,
it is found that the trial states used here and in Ref.~\cite{Yan_2021} can yield accurate dynamics and spectra
at short times, but are less reliable at long times because of augmented accumulative errors in numerics resulting in an increased size of the deviation vector, reminiscence of the situation in a driven, dissipative qubit model~\cite{Yan_2020}.

\begin{figure*}
  \includegraphics[width=2\columnwidth]{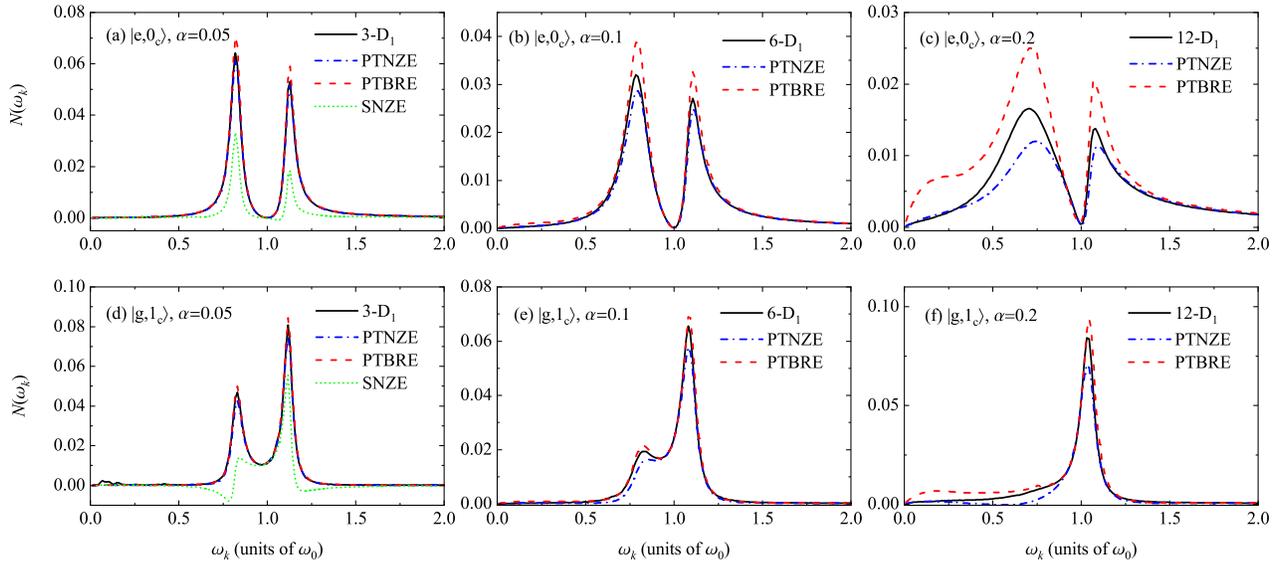}
\caption{Steady-state emission spectra of the dissipative JC model calculated by the variational approach and the master-equation approach
  for $\lambda_c=0.3\omega_0$, $\omega_c=\omega_0$ and the three values of $\alpha$. The qubit-cavity initial states are $|e,0_c\rangle$ and $|g,1_c\rangle$.}\label{fig3}
\end{figure*}

\section{Results and Discussions}\label{sec:RD}

We first examine the performance of the different kinds of non-Markovian quantum regression theories by
calculating the steady-state emission spectrum,
which can be obtained by propagating the equations of motion of the variational approach
and the master equations for a sufficiently long time. Typically, the spectra become stable when $t<200\omega_0^{-1}$ in the ultrastrong coupling regime and thus we show the spectra at $t=200\omega_0^{-1}$ for master equations and variational approach.
In Fig.~\ref{fig2}, we show the steady-state emission spectra of the dissipative Rabi model calculated from PTNZE, PTBRE, SNZE, and the variational methods for $\lambda_c=0.3\omega_0$, $\omega_c=\omega_0$, the three values of $\alpha$, and two kinds of initial states of the qubit-cavity system,
$|e,0_{c}\rangle=|e\rangle\otimes|0_c\rangle$ (i.e., the qubit is in the excited state with the vacuum cavity state) and $|g,1_c\rangle=|g\rangle\otimes|1_c\rangle$ (i.e., the qubit is in the ground state while the cavity is in the single-photon Fock state).
We note that the emission spectra calculated from the PTNZE are satisfactorily in comparison with the numerically exact multi-D$_1$ results for $\alpha\leq0.1$.
Although the deviation between the PTNZE and the multi-D$_1$ results becomes somewhat large for $\alpha=0.2$, the PTNZE treatment remains qualitatively correct.
In addition, we find that in the most cases the PTNZE results deviate less from the multi-D$_1$ than the PTBRE results.
When comparing the emission spectra calculated from SNZE with the multi-D$_1$ results, one notes that the SNZE spectrum is unphysical because of its partial negativity.
Characterizing the distribution of the emitted photon number over the radiation frequency, a physical spectrum is always positive.
Moreover, for $\alpha\geq0.1$, the SNZE yields rather inaccurate results with negative spectral components (which are not shown here).

\begin{figure*}
\includegraphics[width=2\columnwidth]{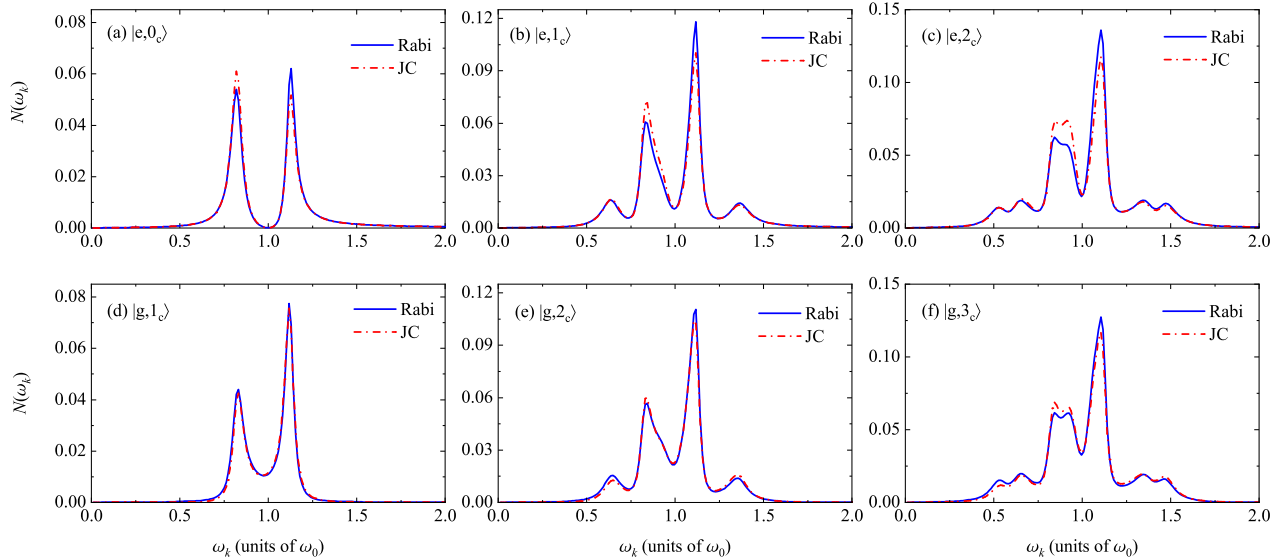}
\caption{Steady-state emission spectra of the dissipative Rabi and JC models calculated by the PTNZE method
for $\lambda_c=0.3\omega_0$, $\omega_c=\omega_0$, $\alpha=0.05$, and the various initial states.}\label{fig4}
\end{figure*}

In Fig.~\ref{fig3}, we show the steady-state emission spectra of the dissipative JC model by using the PTNZE, PTBRE, SNZE, and the variational approach for $\lambda_c=0.3\omega_0$, $\omega_c=\omega_0$, the three values of $\alpha$, and the two kinds of the initial states.
It is found that
the PTNZE provides a quantitatively accurate description of the emission spectrum of the dissipative JC model for $\alpha\leq0.1$ and the deviation between the multi-$D_1$ and PTNZE results are similar as in Fig.~\ref{fig2}.
It should be emphasized that the SNZE breaks down for the three values of $\alpha$.
The present findings confirm the validity of the PTNZE treatment.
Moreover, it is found that in the present problem, the non-Markovian quantum regression theory based on PTNZE is preferred over that based on PTBRE, indicating the importance of the non-Markovianity. In addition,
the PTNZE treatment also performs better than the TRWA treatment in Ref.~\cite{Yan_2021} with the
latter apparently overestimating the high-frequency peak if $\lambda_c>0.1\omega_0$.

We state the computational efficiency of the master-equation-based methods and variational approach when the qubit-cavity system is initially in a single-excitation state and $\lambda_c/\omega_0\sim0.5$. The final propagation time is $t=200\omega_0^{-1}$. It is sufficient to use $N_c=10$, i.e., the number of the cavity Fock states is 10. For the PTNZE, the CPU time is about 4 hours. For the PTBRE, the CPU time is about 1.7 hours. For the multi-D$_1$ method, the CPU time is about 2 hours, 3.2 hours, 7.5 hours, and 41 hours for $M=3$, 4, 6, and 12, respectively. For $M=3$, there are 3060 variational parameters in the ansatz. For $M=12$, there are 12240 variational parameters. It turns out that although a relatively large number of variational parameters is used, the CPU time is still acceptable.

\begin{figure*}
\includegraphics[width=2\columnwidth]{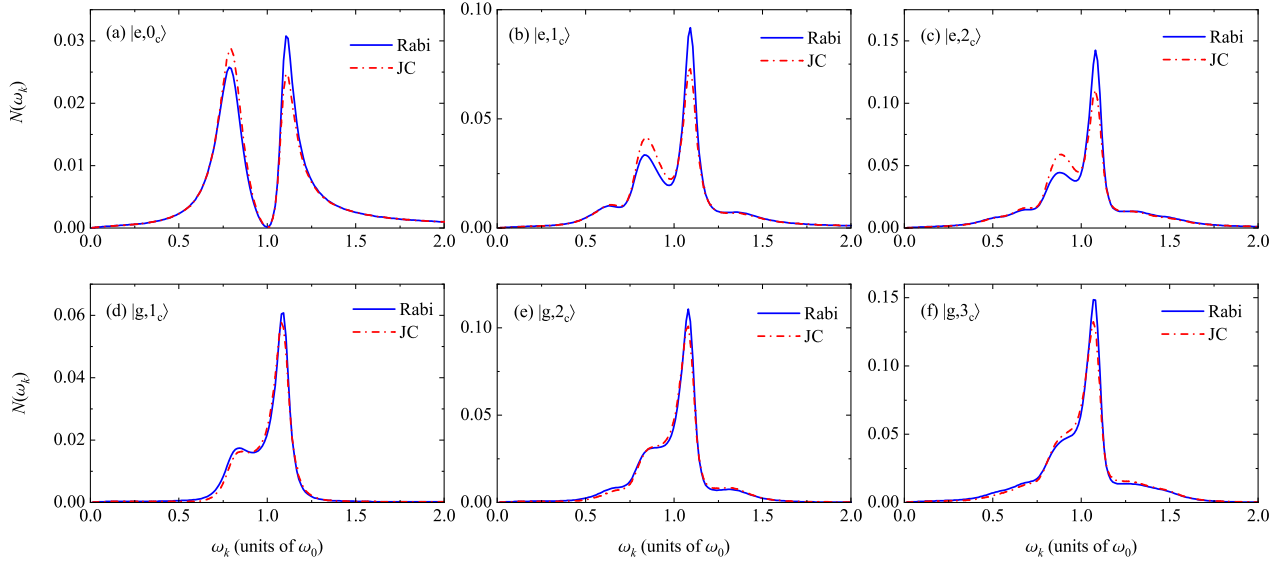}
\caption{Steady-state emission spectra of the dissipative Rabi and JC models calculated
by the PTNZE method for $\lambda_c=0.3\omega_0$, $\omega_c=\omega_0$, $\alpha=0.1$, and the various initial states of the qubit-cavity system.}\label{fig5}
\end{figure*}

\begin{figure*}
\includegraphics[width=2\columnwidth]{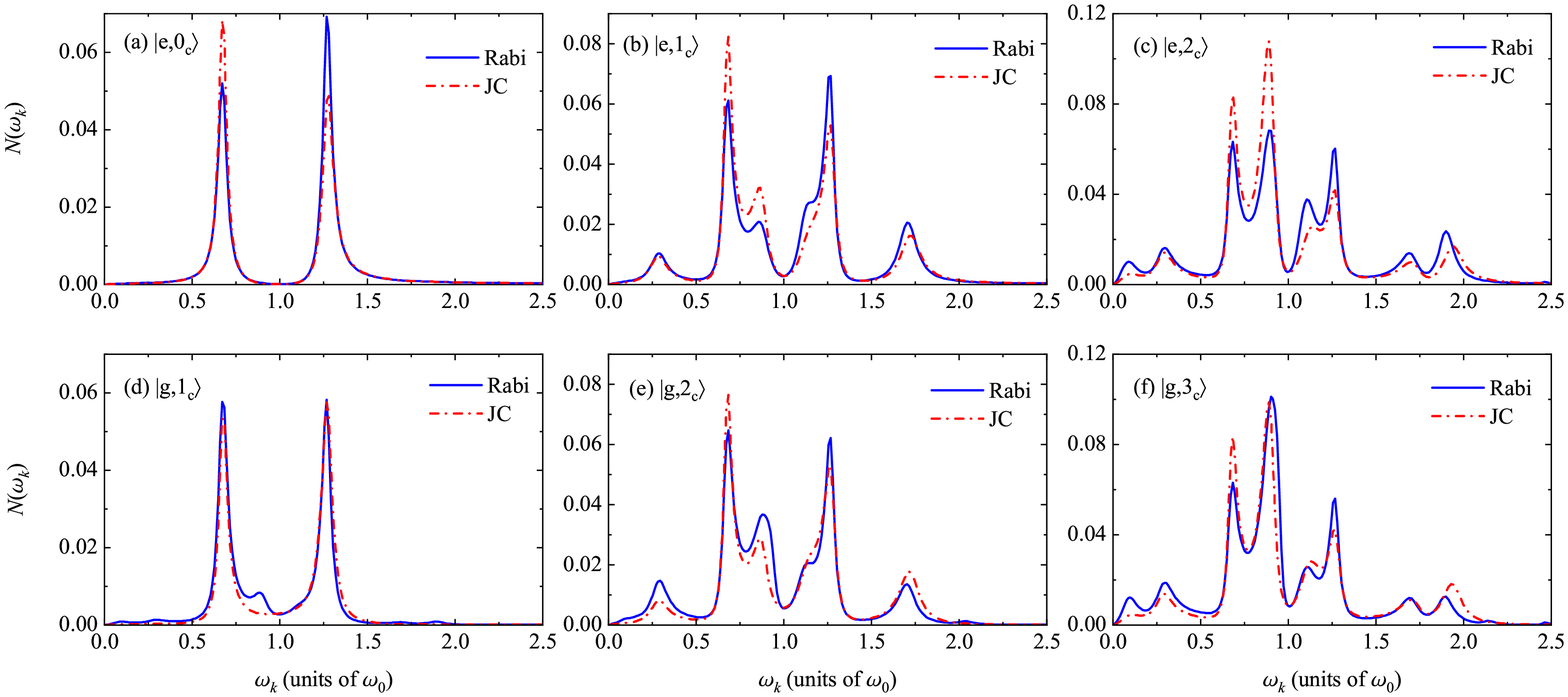}
\caption{Steady-state emission spectra of the dissipative Rabi and JC models calculated by the PTNZE method
for $\lambda_c=0.6\omega_0$, $\omega_c=\omega_0$, $\alpha=0.05$, and the various initial states fo the qubit-cavity system.}\label{fig6}
\end{figure*}

The present PTNZE approach can be applied to the  multiple-excitation state
since the qubit-cavity interaction is treated numerically exactly in this formalism. Employing the PTNZE, we proceed to
study the emission spectra of the dissipative Rabi and JC models under various initial conditions, which helps illustrate the role of the counter-rotating coupling of the qubit-cavity system.

We calculate the emission spectra of the dissipative Rabi model and the dissipative JC model by using the PTNZE method
for $\lambda_c=0.3\omega_0$, $\omega_c=\omega_0$, the two values of $\alpha$, and the various initial states ranging from the single-excitation to the three-excitation initial state.
Figures~\ref{fig4} and \ref{fig5} show the steady-state emission spectra in the cases of $\alpha=0.05$ and $\alpha=0.1$, respectively. Firstly, it is clear to see that in the strong dissipation regime the spectra are generally asymmetric
regardless of whether the counter-rotating terms of the qubit-cavity interaction are taken into account or not.
This is different from the previous finding in the weak-dissipation regime in Ref.~\cite{Cao_2011}, where the RWA spectrum is symmetric while the non-RWA spectrum is asymmetric.
Secondly, one notes that the increase in the excitation number in the initial state leads to that the spectrum is changed from a two-peaked structure to a multipeaked structure in the presence of moderate dissipation.
The formation of the two-peaked and multipeaked spectra can be captured by the dressed-state picture
in which the emission peaks arise from the transition between the dressed states of the qubit-cavity system~\cite{Cohen},
also known as the polariton states.
Particularly, the positions of the peaks can be approximately determined by calculating the energy gaps between the dressed states of the qubit-cavity system participating in the emission processes.
As $\alpha$ increases,
the satellite peaks are likely merged  to form a broad sideband, resulting in an interesting emission line shape.
Thirdly, we note that the deviations between the two models
in Figs.~\ref{fig4}(a)-\ref{fig4}(c) and Figs.~\ref{fig5}(a)-\ref{fig5}(c) are more significant than those in Figs.~\ref{fig4}(d)-\ref{fig4}(f) and Figs.~\ref{fig5}(d)-\ref{fig5}(f), respectively.
This indicates that the influence of the counter-rotating coupling
of the qubit-cavity system depends on the initial states.

To further explore the role of the counter-rotating coupling of the qubit-cavity system,
we calculate the steady-state emission spectra of the dissipative Rabi model and the dissipative JC model by using
the PTNZE for $\lambda_c=0.6\omega_0$, $\omega_c=\omega_0$, $\alpha=0.05$, and various initial states.
With increasing $\lambda_c$, the deviation between the two models
becomes more apparent, revealing several consequences of the counter-rotating terms.
Interestingly, in Figs.~\ref{fig6}(a) and~\ref{fig6}(d), the single-excitation Rabi spectrum is two-peaked for the initial state $|e,0_c\rangle$
while it is multipeaked for the initial state $|g,1_c\rangle$.
Although there is only a single excitation in the initial state $|g,1_c\rangle$, the spectrum is multipeaked
similar to the multiple-excitation spectra.
This is a signature of the nonconversation of the excitation number of the Rabi model due to the counter-rotating coupling.
Actually, for the initial state $|e,0_c\rangle$, the Rabi spectrum also becomes multipeaked when the qubit-cavity coupling is sufficiently strong, e.g., $\lambda_c\sim\omega_0$.
On the contrary, the JC spectrum is two-peaked owing to the absence of the counter-rotating terms of the qubit-cavity interaction,
suggesting that the manifestation of the counter-rotating coupling relies not only on the coupling strength $\lambda_c$ but also on the initial states.
Apart from the additional peaks arising from the counter-rotating terms, those terms can also significantly alter the emission energies and intensities, as shown in Fig.~\ref{fig6}.

\section{Conclusions}\label{sec:Con}
In summary,
we have studied the emission spectra of the Rabi model and the JC model in a dissipative bath in the regime of ultrastrong qubit-cavity coupling, by using the nonMarkvoian quantum regression theory based on PTNZE and the time-dependent variational approach with the multiple Davydov ansatz.
Based on an effective Hamiltonian constructed with a unitary transformation, the PTNZE is found to be satisfactorily accurate as compared to the variational approach for $\alpha\leq0.1$, a coupling regime that is beyond the capability of SNZE. Using the PTNZE, the role played by the qubit-cavity counter-rotating term on the emission spectrum has been illustrated by spectral comparison of the two models. It is shown that the emission spectrum is generally asymmetric
regardless of the qubit-cavity counter-rotating term, a result at variance with the findings in the weak-dissipation limit~\cite{Cao_2011}.
For single-excitation initialization, the time-dependent variational approach with the multiple Davydov ansatz is capable to yield accurate steady-state behavior, and it is found that the counter-rotating term leads to excitation nonconservation and multipeaked emission spectra. The effect of the counter-rotating term on emission spectra, however, hinges on the coupling strength and the system initialization. For multiple-excitation initial states, improvement of the multiple Davydov ansatz may be needed to better describe steady states.

The present approaches can be applied to the problems ranging from cavity-QED to molecular vibrational polaritons~\cite{Xiang_2021,Engelhardt2022} or light-harvesting complexes described by the  Fenna-Mathews-Olson model.

\begin{acknowledgements}
Support from the National Natural Science Foundation of China (Grants No. 12005188, No. 11774226, and No. 11774311)
and the Singapore Ministry of Education Academic Research
Fund Tier 1 (Grant No. RG87/20) is gratefully acknowledged.
\end{acknowledgements}

\appendix
\section{Effective Hamiltonian and master equations for the dissipative Rabi model}\label{app:me}
The unitary transformation can be done straightforwardly and the resulting Hamiltonian
can be partitioned into the three parts~\cite{Zheng_2004}:
\begin{equation}
H^{\prime}=H_{0}^{\prime}+H_{1}^{\prime}+H_{2}^{\prime},
\end{equation}
\begin{eqnarray}
H_{0}^{\prime} & = & \frac{1}{2}\eta\omega_{0}\sigma_{z}+\omega_{c}\hat{b}_{c}^{\dagger}\hat{b}_{c}+\frac{\lambda_{c}}{2}(\hat{b}_{c}^{\dagger}+\hat{b}_{c})\sigma_{x}+\sum_{k}\omega_{k}\hat{b}_{k}^{\dagger}\hat{b}_{k}\nonumber \\
 &  & -\sum_{k}\frac{\lambda_{k}^{2}}{4\omega_{k}}\xi_{k}(2-\xi_{k}),
\end{eqnarray}
\begin{eqnarray}
H_{1}^{\prime} & = & \sum_{k}\frac{\lambda_{k}}{2}(1-\xi_{k})(\hat{b}_{k}^{\dagger}+\hat{b}_{k})\sigma_{x}\nonumber\\
&  &-\frac{i\eta\omega_{0}}{2}\sum_{k}\frac{\lambda_{k}}{\omega_{k}}\xi_{k}(\hat{b}_{k}^{\dagger}-\hat{b}_{k})\sigma_{y},
\end{eqnarray}
\begin{eqnarray}
H_{2}^{\prime} & = & \frac{\omega_{0}}{2}\sigma_{z}\left(\cosh X-\eta\right)-i\frac{\omega_{0}}{2}\sigma_{y}\left(\sinh X-\eta X\right),
\end{eqnarray}
where
\begin{equation}
  X=\sum_{k}\frac{\lambda_{k}}{\omega_{k}}\xi_{k}(\hat{b}_{k}^{\dagger}-\hat{b}_{k})
\end{equation}
\begin{equation}
\eta=\langle\left\{ 0_{k}\right\} |\cosh X|\left\{ 0_{k}\right\} \rangle=\exp\left[-\sum_{k}\frac{1}{2}\left(\frac{\lambda_{k}}{\omega_{k}}\xi_{k}\right)^{2}\right].
\end{equation}
The parameter $\xi_{k}$ is determined by equating the coefficients in $H_1^\prime$, i.e.,
\begin{equation}
\frac{\lambda_{k}}{2}(1-\xi_{k})=\frac{1}{2}\eta\omega_{0}\frac{\lambda_{k}}{\omega_{k}}\xi_{k},\label{eq:xik}
\end{equation}
which yields
\begin{equation}
\xi_{k}=\frac{\omega_{k}}{\eta\omega_{0}+\omega_{k}}.
\end{equation}
As a result, $H_{1}^{\prime}$ takes the rotating-wave form, i.e.,
\begin{equation}
H_{1}^{\prime}=\sum_{k}\tilde{\lambda}_{k}(\hat{b}_{k}^{\dagger}\sigma_{-}+\hat{b}_{k}\sigma_{+}),
\end{equation}
where the modified coupling constants are
\begin{equation}
\tilde{\lambda}_{k}=\frac{\eta\omega_{0}\lambda_{k}}{\eta\omega_{0}+\omega_{k}}.
\end{equation}

In $H_0^\prime$, the qubit-cavity system is decoupled from the reservoir. $H_1^\prime$ describes the modified first-order interaction between the qubit and the reservoir. $H_2^\prime$ is the higher-order interaction, which can be safely dropped under certain conditions. Therefore, we use $H^\prime\approx H_0^\prime+H_1^\prime$ as the effective Hamiltonian.

In the transformed frame, we have
\begin{eqnarray}
\Lambda^{\prime}(t,t^{\prime}) & = & e^{S}\Lambda(t,t^{\prime})e^{-S}\nonumber\\
 & = & e^{S}U(t-t^{\prime})\sigma_{x}\rho(t^{\prime})U^{\dagger}(t-t^{\prime})e^{-S}\nonumber\\
 & = & e^{S}U(t-t^{\prime})e^{-S}\sigma_{x}e^{S}\rho(t^{\prime})e^{-S}e^{S}U^{\dagger}(t-t^{\prime})e^{-S}\nonumber\\
 & = & e^{-iH^{\prime}(t-t^{\prime})}\sigma_{x}\rho^{\prime}(t^{\prime})e^{iH^{\prime}(t-t^{\prime})}.
\end{eqnarray}
We  set
$H_{0}^{\prime}$ and $H_{1}^{\prime}$ as the free and interaction
Hamiltonian, respectively. We first derive the equation of motion
for the reduced operator $\Lambda_{{\rm S}}^{\prime}(t,t^{\prime})={\rm Tr}_{\rm R}\Lambda^{\prime}(t,t^{\prime})$.
In the interaction picture, the operator $\Lambda^{\prime I}(t,t^{\prime})=\exp(iH_{0}^{\prime}t)\Lambda^{\prime}(t,t^{\prime})\exp(-iH_{0}^{\prime}t)$
satisfies
\begin{equation}
\frac{d}{dt}\Lambda^{\prime I}(t,t^{\prime})=-i[H_{1}^{\prime}(t),\Lambda^{\prime I}(t,t^{\prime})],\label{eq:VNE}
\end{equation}
where
\begin{equation}
H_{1}^{\prime}(t)=\sum_{k}\tilde{\lambda}_{k}[\hat{b}_{k}^{\dagger}e^{i\omega_{k}t}\sigma_{-}(t)+\hat{b}_{k}e^{-i\omega_{k}t}\sigma_{+}(t)],
\end{equation}
\begin{equation}
\sigma_{\pm}(t)=e^{iH_{{\rm Rabi}}^{\prime}t}\sigma_{\pm}e^{-iH_{{\rm Rabi}}^{\prime}t}.
\end{equation}
By formal integral, one finds
\begin{equation}
\Lambda^{\prime I}(t,t^{\prime})=\Lambda^{\prime I}(t^{\prime},t^{\prime})-i\int_{t^{\prime}}^{t}d\tau[H_{1}^{\prime}(\tau),\Lambda^{\prime I}(\tau,t^{\prime})].\label{eq:formalsol}
\end{equation}
Substituting the formal solution into Eq. (\ref{eq:VNE}), one has
\begin{eqnarray}
\frac{d}{dt}\Lambda^{\prime I}(t,t^{\prime})&=&-i[H_{1}^{\prime}(t),\sigma_{x}(t^{\prime})\rho^{I}(t^{\prime})]\nonumber\\
&  &-\int_{t^{\prime}}^{t}d\tau[H_{1}^{\prime}(t),[H_{1}^{\prime}(\tau),\Lambda^{\prime I}(\tau,t^{\prime})]],\label{eq:LA1}
\end{eqnarray}
where we have used $\Lambda^{\prime I}(t^{\prime},t^{\prime})=\sigma_{x}(t^{\prime})\rho^{\prime I}(t^{\prime})$.
Here $\rho^{\prime I}(t^{\prime})$ is the density matrix of the total
system at time $t^{\prime},$which satisfies
\begin{equation}
\frac{d}{dt}\rho^{\prime I}(t^{\prime})=-i[H_{1}^{\prime}(t^{\prime}),\rho^{\prime I}(t^{\prime})].
\end{equation}
The total density matrix takes on the formal solution
\begin{equation}
\rho^{\prime I}(t^{\prime})=\rho^{\prime I}(0)-i\int_{0}^{t^{\prime}}d\tau[H_{1}^{\prime}(\tau),\rho^{\prime I}(\tau)].
\end{equation}
Substituting the above equation into Eq.~(\ref{eq:LA1}), we arrive at
\begin{eqnarray}
\frac{d}{dt}\Lambda^{\prime I}(t,t^{\prime})&=&-i[H_{I}^{\prime}(t),\sigma_{x}(t^{\prime})\rho^{\prime I}(0)]\nonumber\\
&  &-\int_{0}^{t^{\prime}}d\tau[H_{1}^{\prime}(t),\sigma_{x}(t^{\prime})[H_{1}^{\prime}(\tau),\rho^{\prime I}(\tau)]\nonumber\\
&  &-\int_{t^{\prime}}^{t}d\tau[H_{1}^{\prime}(t),[H_{1}^{\prime}(\tau),\Lambda^{\prime I}(\tau,t^{\prime})]],\label{eq:LA2}
\end{eqnarray}
To proceed, we use the Born approximations $\Lambda^{\prime I}(\tau,t^{\prime})\approx\Lambda_{{\rm S}}^{\prime I}(\tau,t^{\prime})|\left\{ 0_{k}\right\} \rangle\langle\left\{ 0_{k}\right\} |$
and $\rho^{\prime I}(\tau)\approx\rho_{{\rm S}}^{\prime I}(\tau)|\left\{ 0_{k}\right\} \rangle\langle\left\{ 0_{k}\right\} |$
and assume a factorized initial condition $\rho^{\prime I}(0)=\rho_{{\rm S}}^{\prime}(0)|\left\{ 0_{k}\right\} \rangle\langle\left\{ 0_{k}\right\} |$.
Taking the partial trace over the reservoir for Eq.~(\ref{eq:LA2})
and transforming into the Schr\"{o}dinger picture, one readily derives
the equation of motion for the reduced operator.
Clearly, a similar procedure leads to the PTNZE
of the reduced density matrix.

The master equations are differential-integral equations, which can be solved
by the following method. For concrete derivation, we take Eq.~(\ref{eq:tnlme2})
as an example. We first fit the bath correlation function with the
exponential functions~\cite{Duan_2017}
\begin{equation}
\tilde{C}(t)=\sum_{l=1}^{K}g_{l}e^{-\gamma_{l}t},
\end{equation}
where $g_{l}$ and $\gamma_{l}$ (${\rm Re}\gamma_{l}>0$) are the
complex-valued parameters. Second, we define the auxiliary density
matrix as follows:
\begin{equation}
\rho_{l}(t)=\int_{0}^{t}d\tau g_{l}e^{-\gamma_{l}(t-\tau)}e^{-iH_{{\rm Rabi}}^{\prime}(t-\tau)}\sigma_{-}\rho_{{\rm S}}^{\prime}(\tau)e^{iH_{{\rm Rabi}}^{\prime}(t-\tau)}.
\end{equation}
Taking the derivative with respect to $t$, one finds the equation
of motion for auxiliary density matrix
\begin{equation}
\frac{d}{dt}\rho_{l}(t)=-i[H_{{\rm Rabi}}^{\prime},\rho_{l}(t)]-\gamma_{l}\rho_{l}(t)+g_{l}\sigma_{-}\rho_{S}^{\prime}(t),\label{eq:auxirho}
\end{equation}
with $\rho_{l}(0)$ being the zero matrix. In doing so, Eq. (\ref{eq:tnlme2})
becomes
\begin{equation}
\frac{d}{dt}\rho_{{\rm S}}^{\prime}(t)=-i[H_{{\rm Rabi}}^{\prime},\rho_{S}^{\prime}(t)]-\left\{ [\sigma_{+},\sum_{l=1}^{K}\rho_{l}(t)]+{\rm h.c.}\right\} .\label{eq:tnlmefinal}
\end{equation}
We can combine Eq. (\ref{eq:auxirho}) together with (\ref{eq:tnlmefinal})
and solve them with the Runge-Kutta method. Clearly, the SNZE presented in the main text can also be solved by the above
method.
\section{Effective Hamiltonian and master equations of the dissipative JC model}\label{app:JCHeff}
To transform the dissipative JC Hamiltonian, we rewrite $H_{\rm JC}$ in an alternative form:
\begin{equation}
H_{{\rm JC}}=\frac{1}{2}\omega_{0}\sigma_{z}+\omega_{c}\hat{b}_{c}^{\dagger}\hat{b}_{c}+\frac{\lambda_{c}}{4}(\hat{b}_{c}+\hat{b}_{c}^{\dagger})\sigma_{x}+i\frac{\lambda_{c}}{4}(\hat{b}_{c}-\hat{b}_{c}^{\dagger})\sigma_{y}.
\end{equation}
The unitary transformation can be applied to the total Hamiltonian as follows:
\begin{equation}
H^{\prime}=e^{S}(H_{{\rm JC}}+H_{{\rm R}}+H_{{\rm I}})e^{-S}=H_{0}^{\prime}+H_{1}^{\prime}+H_{2}^{\prime},
\end{equation}
\begin{equation}
H_{0}^{\prime}=H_{{\rm JC}}^{\prime}+\sum_{k}\omega_{k}\hat{b}_{k}^{\dagger}\hat{b}_{k}-\sum_{k}\frac{\lambda_{k}^{2}}{4\omega_{k}}\xi_{k}(2-\xi_{k}),
\end{equation}
\begin{eqnarray}
H_{1}^{\prime} & = & \sum_{k}\frac{\lambda_{k}}{2}(1-\xi_{k})(\hat{b}_{k}^{\dagger}+\hat{b}_{k})\sigma_{x}\nonumber\\
&  &-\frac{i\eta\omega_{0}}{2}\sum_{k}\frac{\lambda_{k}}{\omega_{k}}\xi_{k}(\hat{b}_{k}^{\dagger}-\hat{b}_{k})\sigma_{y}\nonumber \\
 &  & -\eta\frac{\lambda_{c}}{4}\sum_{k}\frac{\lambda_{k}}{\omega_{k}}\xi_{k}(\hat{b}_{k}^{\dagger}-\hat{b}_{k})(\hat{b}_{c}-\hat{b}_{c}^{\dagger})\sigma_{z},
\end{eqnarray}

\begin{eqnarray}
H_{2}^{\prime} & = & \left[i\frac{\omega_{0}}{2}\sigma_{y}+\frac{\lambda_{c}}{4}(\hat{b}_{c}-\hat{b}_{c}^{\dagger})\sigma_{z}\right](\eta X-\sinh X)\nonumber \\
 &  &+\left[\frac{\omega_{0}}{2}\sigma_{z}+i\frac{\lambda_{c}}{4}(\hat{b}_{c}-\hat{b}_{c}^{\dagger})\sigma_{y}\right](\cosh X-\eta).
\end{eqnarray}
where
\begin{equation}
H_{{\rm JC}}^{\prime}=\frac{1}{2}\eta\omega_{0}\sigma_{z}+\omega_{c}\hat{b}_{c}^{\dagger}\hat{b}_{c}+\frac{\lambda_{c}}{4}(\hat{b}_{c}+\hat{b}_{c}^{\dagger})\sigma_{x}+i\frac{\eta\lambda_{c}}{4}(\hat{b}_{c}-\hat{b}_{c}^{\dagger})\sigma_{y}.
\end{equation}
To proceed, the parameter $\xi_{k}$ is obtained by using Eq.~\eqref{eq:xik}.
In doing so, $H_{1}^{\prime}$ is rewritten as follows:
\begin{eqnarray}
H_{1}^{\prime} & = & \sum_{k}\tilde{\lambda}_{k}\left\{ \hat{b}_{k}^{\dagger}\left[\sigma_{-}-\frac{\lambda_{c}}{4\omega_{0}}(\hat{b}_{c}-\hat{b}_{c}^{\dagger})\sigma_{z}\right]\right.\nonumber\\
& &\left.+\hat{b}_{k}\left[\sigma_{+}+\frac{\lambda_{c}}{4\omega_{0}}(\hat{b}_{c}-\hat{b}_{c}^{\dagger})\sigma_{z}\right]\right\} \nonumber \\
 & \equiv & \sum_{k}\tilde{\lambda}_{k}(\hat{b}_{k}^{\dagger}\hat{S}_{-}+\hat{b}_{k}\hat{S}_{+}),
\end{eqnarray}
where
\begin{equation}
\hat{S}_{\pm}=\sigma_{\pm}\pm\frac{\lambda_{c}}{4\omega_{0}}(\hat{b}_{c}-\hat{b}_{c}^{\dagger})\sigma_{z}.
\end{equation}
The effective Hamiltonian is set as $H^\prime\approx H^\prime_0+H^\prime_1$. The master equations for the dissipative JC model can be obtained
directly from Eqs. \eqref{eq:tnlme1} and \eqref{eq:tnlme2} by the following substitutions:
\begin{equation}
H_{{\rm Rabi}}^{\prime}\rightarrow H_{{\rm JC}}^{\prime},
\end{equation}
\begin{equation}
\sigma_{\pm}\rightarrow\hat{S}_{\pm}.
\end{equation}

\section{The equations of motion for the variational parameters}\label{app:eom}
By substituting the ansatz into Eq.~(\ref{eq:tdvp}) in the main text, one finds the equations of motion,
which reads
\begin{widetext}
 \begin{equation}
i\langle+|\langle\phi_{j}|\langle f_{m}|\dot{D}^{M}_{1}(t)\rangle=\langle+|\langle\phi_{j}|\langle f_{m}|\tilde{H}(t)|D^{M}_{1}(t)\rangle,\label{eq:eom1}
\end{equation}
\begin{equation}
i\sum_{j=1}^{N_c}A_{mj}^{\ast}\langle+|\langle\phi_{j}|\langle f_{m}|\hat{b}_{p}|\dot{D}^{M}_{1}(t)\rangle=\sum_{j=1}^{N_c}A_{mj}^{\ast}\langle+|\langle\phi_{j}|\langle f_{m}|\hat{b}_{p}\tilde{H}(t)|D^{M}_{1}(t)\rangle,
\end{equation}
\begin{equation}
i\langle-|\langle\phi_{j}|\langle g_{m}|\dot{D}^{M}_{1}(t)\rangle=\langle-|\langle\phi_{j}|\langle g_{m}|\tilde{H}(t)|D^{M}_{1}(t)\rangle,
\end{equation}
\begin{equation}
i\sum_{j=1}^{N_c}B_{mj}^{\ast}\langle-|\langle\phi_{j}|\langle g_{m}|\hat{b}_{p}|\dot{D}^{M}_{1}(t)\rangle=\sum_{j=1}^{N_c}B_{mj}^{\ast}\langle-|\langle\phi_{j}|\langle g_{m}|\hat{b}_{p}\tilde{H}(t)|D^{M}_{1}(t)\rangle.\label{eq:eom4}
\end{equation}
\end{widetext}

The time derivative of the trial state is given by
\begin{eqnarray}
|\dot{D}^{M}_{1}(t)\rangle & = & \sum_{d=1}^{M}\sum_{i=1}^{N_c}a_{di}|+\rangle|\phi_{i}\rangle|f_{d}\rangle\nonumber\\
&  &+\sum_{d=1}^{M}\sum_{i=1}^{N_c}A_{di}\sum_{k=1}^{N_b}\dot{f}_{dk}\hat{b}_{k}^{\dagger}|+\rangle|\phi_{i}\rangle|f_{d}\rangle\nonumber \\
 &  & +\sum_{d=1}^{M}\sum_{i=1}^{N_c}b_{di}|-\rangle|\phi_{i}\rangle|g_{d}\rangle\nonumber\\
 &  &+\sum_{d=1}^{M}\sum_{i=1}^{N_c}B_{di}\sum_{k=1}^{N_b}\dot{g}_{dk}\hat{b}_{k}^{\dagger}|-\rangle|\phi_{i}\rangle|g_{d}\rangle.
\end{eqnarray}
where
\begin{equation}
a_{di}=\dot{A}_{di}-\frac{1}{2}A_{di} \sum_{k=1}^{N_b}\left(f_{dk}\dot{f}_{dk}^{\ast}+\dot{f}_{dk}f_{dk}^{\ast}\right),\label{eq:adi}
\end{equation}

\begin{equation}
b_{di}=\dot{B}_{di}-\frac{1}{2}B_{di}\sum_{k=1}^{N_b}\left(g_{dk}\dot{g}_{dk}^{\ast}+\dot{g}_{dk}g_{dk}^{\ast}\right).\label{eq:bdi}
\end{equation}

\begin{figure*}
  \centering
  \includegraphics[width=1.2\columnwidth]{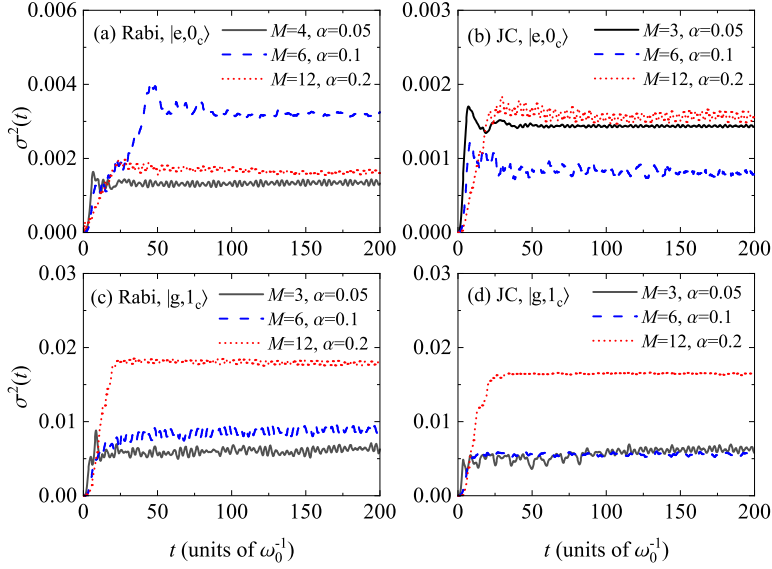}
  \caption{Squared norm of the deviation vector $\sigma^2(t)$ versus time $t$ for the variational results in Figs.~\ref{fig2} and \ref{fig3}.}\label{fig7}
\end{figure*}

The inner products on the left hand side of Eqs. (\ref{eq:eom1})-(\ref{eq:eom4})
can thus be calculated as follows:
\begin{widetext}
\begin{equation}
\langle+|\langle\phi_{j}|\langle f_{m}|\dot{D}^{M}_{1}(t)\rangle=\sum_{d=1}^{M}\left(a_{dj}+A_{dj}\sum_{k=1}^{N_b}f_{mk}^{\ast}\dot{f}_{dk}\right)\langle f_{m}|f_{d}\rangle,\label{eq:td1}
\end{equation}
\begin{equation}
\sum_{j=1}^{N_c}A_{mj}^{\ast}\langle+|\langle\phi_{j}|\langle f_{m}|\hat{b}_{p}|\dot{D}^{M}_{1}(t)\rangle=\sum_{d=1}^{M}\left(\sum_{j=1}^{N_c}A_{mj}^{\ast}a_{dj}f_{dp}+\sum_{j=1}^{N_c}A_{mj}^{\ast}A_{dj}\sum_{k=1}^{N_b}(\delta_{k,p}+f_{mk}^{\ast}f_{dp})\dot{f}_{dk}\right)\langle f_{m}|f_{d}\rangle,
\end{equation}
\begin{equation}
\langle-|\langle\phi_{j}|\langle g_{m}|\dot{D}^{M}_{1}(t)\rangle=\sum_{d=1}^{M}\left(b_{dj}+B_{dj}\sum_{k=1}^{N_b}g_{mk}^{\ast}\dot{g}_{dk}\right)\langle g_{m}|g_{d}\rangle,
\end{equation}
\begin{equation}
\sum_{j=1}^{N_c}B_{mj}^{\ast}\langle-|\langle\phi_{j}|\langle g_{m}|\hat{b}_{p}|\dot{D}^{M}_{1}(t)\rangle=\sum_{d=1}^{M}\left(\sum_{j=1}^{N_c}B_{mj}^{\ast}\mathit{b}_{dj}g_{dp}+\sum_{j=1}^{N_c}B_{mj}^{\ast}B_{dj}\sum_{k=1}^{N_b}(\delta_{k,p}+g_{mk}^{\ast}g_{dp})\dot{g}_{dk}\right)\langle g_{m}|g_{d}\rangle.\label{eq:td2}
\end{equation}
The imhomogeneous terms on the right hand side of Eqs. (\ref{eq:eom1})-(\ref{eq:eom4})
are given by
\begin{eqnarray}
\langle+|\langle\phi_{j}|\langle f_{m}|\tilde{H}(t)|D^{M}_{1}(t)\rangle & = & \sum_{d=1}^{M}\sum_{i=1}^{N_c}\left(A_{di}\langle+|\langle\phi_{j}|H_{{\rm Rabi}}|+\rangle|\phi_{i}\rangle\langle f_{m}|f_{d}\rangle+B_{di}\langle+|\langle\phi_{j}|H_{{\rm Rabi}}|-\rangle|\phi_{i}\rangle\langle f_{m}|g_{d}\rangle\right)\nonumber \\
 &  & +\sum_{d=1}^{M}A_{dj}\sum_{k=1}^{N_b}\frac{\lambda_{k}}{2}(f_{mk}^{\ast}e^{i\omega_{k}t}+f_{dk}e^{-i\omega_{k}t})\langle f_{m}|f_{d}\rangle,\label{eq:vecI1}
\end{eqnarray}
\begin{eqnarray}
\sum_{j=1}^{N_c}A_{mj}^{\ast}\langle+|\langle\phi_{j}|\langle f_{m}|\hat{b}_{p}\tilde{H}(t)|D^{M}_{1}(t)\rangle & = & \sum_{d=1}^{M}\sum_{j,i=1}^{N_c}\left(A_{mj}^{\ast}A_{di}\langle+|\langle\phi_{j}|H_{{\rm Rabi}}|+\rangle|\phi_{i}\rangle f_{dp}\langle f_{m}|f_{d}\rangle\right.\nonumber\\
&  &\left.+A_{mj}^{\ast}B_{di}\langle+|\langle\phi_{j}|H_{{\rm Rabi}}|-\rangle|\phi_{i}\rangle g_{dp}\langle f_{m}|g_{d}\rangle\right)\nonumber \\
 &  & +\sum_{d=1}^{M}\sum_{j=1}^{N_c}A_{mj}^{\ast}A_{dj}\sum_{k=1}^{N_b}\frac{\lambda_{k}}{2}[\delta_{k,p}e^{i\omega_{k}t}+(f_{mk}^{\ast}e^{i\omega_{k}t}+f_{dk}e^{-i\omega_{k}t})f_{dp}]\langle f_{m}|f_{d}\rangle,
\end{eqnarray}
\begin{eqnarray}
\langle-|\langle\phi_{j}|\langle g_{m}|\tilde{H}(t)|D^{M}_{1}(t)\rangle & = & \sum_{d=1}^{M}\sum_{i=1}^{N_c}\left(A_{di}\langle-|\langle\phi_{j}|H_{{\rm Rabi}}|+\rangle|\phi_{i}\rangle\langle g_{m}|f_{d}\rangle+B_{di}\langle-|\langle\phi_{j}|H_{{\rm Rabi}}|-\rangle|\phi_{i}\rangle\langle g_{m}|g_{d}\rangle\right)\nonumber \\
 &  & -\sum_{d=1}^{M}B_{dj}\sum_{k=1}^{N_b}\frac{\lambda_{k}}{2}(g_{mk}^{\ast}e^{i\omega_{k}t}+g_{dk}e^{-i\omega_{k}t})\langle g_{m}|g_{d}\rangle,
\end{eqnarray}
\begin{eqnarray}
\sum_{j=1}^{N_c}B_{mj}^{\ast}\langle-|\langle\phi_{j}|\langle g_{m}|\hat{b}_{p}\tilde{H}(t)|D^{M}_{1}(t)\rangle & = & \sum_{d=1}^{M}\sum_{j,i=1}^{N_c}\left(B_{mj}^{\ast}A_{di}\langle+|\langle\phi_{j}|H_{{\rm Rabi}}|+\rangle|\phi_{i}\rangle f_{dp}\langle g_{m}|f_{d}\rangle\right.\nonumber\\
&  & \left.+B_{mj}^{\ast}B_{di}\langle-|\langle\phi_{j}|H_{{\rm Rabi}}|-\rangle|\phi_{i}\rangle g_{dp}\langle g_{m}|g_{d}\rangle\right)\nonumber \\
 &  & -\sum_{d=1}^{M}\sum_{j=1}^{N_c}B_{mj}^{\ast}B_{dj}\sum_{k=1}^{N_b}\frac{\lambda_{k}}{2}[\delta_{k,p}e^{i\omega_{k}t}+(g_{mk}^{\ast}e^{i\omega_{k}t}+g_{dk}e^{-i\omega_{k}t})g_{dp}]\langle g_{m}|g_{d}\rangle.\label{eq:vecI4}
\end{eqnarray}

To perform the simulation, the equations of motion are rewritten in
a matrix form $i{\cal M}\vec{y}=\vec{I}$, where ${\cal M}$ is the
coefficient matrix defined in Eqs. (\ref{eq:td1})-(\ref{eq:td2}),
 the vector $\vec{y}$ is composed of the quantities $\{a_{di},\dot{f}_{dk},b_{di},\dot{g}_{dk}\}$,
and $\vec{I}$ is the imhomogeneous vector the components of which are
defined in Eqs. (\ref{eq:vecI1})-(\ref{eq:vecI4}). To obtain the
time derivative of the variational parameters, one should solve the
matrix equation for $\vec{y}$. This yields the values for the variables
$\{a_{di},\dot{f}_{dk},b_{di},\dot{g}_{dk}\}$. By using Eqs. (\ref{eq:adi})
and (\ref{eq:bdi}), one calculates the time derivative of $A_{di}$
and $B_{di}$. On obtaining the time derivative of the variational
parameters, we use the $4$th-order Runge-Kutta method to update the
values of the variational parameters.

To measure the accuracy of the variational results, we need the
mean value of $\tilde{H}^{2}(t)$ and the squared norm of $|\dot{D}^{M}(t)\rangle$,
which can be calculated directly as follows:
\begin{eqnarray}
\left\langle \tilde{H}^{2}(t)\right\rangle  & = & \sum_{m,d=1}^{M}\sum_{j,i=1}^{N_c}\left(A_{mj}^{\ast}A_{di}\langle+|\langle\phi_{j}|H_{{\rm Rabi}}^{2}|+\rangle|\phi_{i}\rangle\langle f_{m}|f_{d}\rangle+A_{mj}^{\ast}B_{di}\langle+|\langle\phi_{j}|H_{{\rm Rabi}}^{2}|-\rangle|\phi_{i}\rangle\langle f_{m}|g_{d}\rangle\right.\nonumber \\
 &  & \left.+B_{mj}^{\ast}A_{di}\langle-|\langle\phi_{j}|H_{{\rm Rabi}}^{2}|+\rangle|\phi_{i}\rangle\langle g_{m}|f_{d}\rangle+B_{mj}^{\ast}B_{di}\langle-|\langle\phi_{j}|H_{{\rm Rabi}}^{2}|-\rangle|\phi_{i}\rangle\langle g_{m}|g_{d}\rangle\right)\nonumber \\
 &  & +\sum_{m,d=1}^{M}\sum_{j,i=1}^{N_c}\left(A_{mj}^{\ast}A_{di}\langle+|\langle\phi_{j}|\left\{ H_{{\rm Rabi}},\sigma_{x}\right\} |+\rangle|\phi_{i}\rangle\sum_{k=1}^{N_b}\frac{\lambda_{k}}{2}(f_{mk}^{\ast}e^{i\omega_{k}t}+f_{dk}e^{-i\omega_{k}t})\langle f_{m}|f_{d}\rangle\right.\nonumber \\
 &  & +A_{mj}^{\ast}B_{di}\langle+|\langle\phi_{j}|\left\{ H_{{\rm Rabi}},\sigma_{x}\right\} |-\rangle|\phi_{i}\rangle\sum_{k=1}^{N_b}\frac{\lambda_{k}}{2}(f_{mk}^{\ast}e^{i\omega_{k}t}+g_{dk}e^{-i\omega_{k}t})\langle f_{m}|g_{d}\rangle\nonumber \\
 &  & +B_{mj}^{\ast}A_{di}\langle-|\langle\phi_{j}|\left\{ H_{{\rm Rabi}},\sigma_{x}\right\} |+\rangle|\phi_{i}\rangle\sum_{k=1}^{N_b}\frac{\lambda_{k}}{2}(g_{mk}^{\ast}e^{i\omega_{k}t}+f_{dk}e^{-i\omega_{k}t})\langle g_{m}|f_{d}\rangle\nonumber \\
 &  & \left.+B_{mj}^{\ast}B_{di}\langle-|\langle\phi_{j}|\left\{ H_{{\rm Rabi}},\sigma_{x}\right\} |-\rangle|\phi_{i}\rangle\sum_{k=1}^{N_b}\frac{\lambda_{k}}{2}(g_{mk}^{\ast}e^{i\omega_{k}t}+g_{dk}e^{-i\omega_{k}t})\langle g_{m}|g_{d}\rangle\right)\nonumber \\
 &  & +\sum_{m,d=1}^{M}\sum_{i=1}^{N_c}A_{mi}^{\ast}A_{di}\left\{ \left[\sum_{k=1}^{N_b}\frac{\lambda_{k}}{2}(f_{mk}^{\ast}e^{i\omega_{k}t}+f_{dk}e^{-i\omega_{k}t})\right]^{2}+\sum_{k=1}^{N_b}\frac{\lambda_{k}^{2}}{4}\right\} \langle f_{m}|f_{d}\rangle\nonumber \\
 &  & +\sum_{m,d=1}^{M}\sum_{i=1}^{N_c}B_{mi}^{\ast}B_{di}\left\{ \left[\sum_{k=1}^{N_b}\frac{\lambda_{k}}{2}(g_{mk}^{\ast}e^{i\omega_{k}t}+g_{dk}e^{-i\omega_{k}t})\right]^{2}+\sum_{k=1}^{N_b}\frac{\lambda_{k}^{2}}{4}\right\} \langle g_{m}|g_{d}\rangle,
\end{eqnarray}
\begin{eqnarray}
\langle\dot{D}^{M}_{1}(t)|\dot{D}^{M}_{1}(t)\rangle & = & \sum_{m,d=1}^{M}\sum_{i=1}^{N_c}\left(a_{mi}^{\ast}a_{di}+a_{mi}^{\ast}A_{di}\sum_{k=1}^{N_b}f_{mk}^{\ast}\dot{f}_{dk}+A_{mi}^{\ast}a_{di}\sum_{k=1}^{N_b}\dot{f}_{mk}^{\ast}f_{dk}\right)\langle f_{m}|f_{d}\rangle\nonumber \\
 &  & +\sum_{m,d=1}^{M}\sum_{i=1}^{N_c}A_{mi}^{\ast}A_{di}\sum_{k,p=1}^{N_b}(\delta_{k,p}+f_{mk}^{\ast}f_{dp})\dot{f}_{mp}^{\ast}\dot{f}_{dp}\langle f_{m}|f_{d}\rangle\nonumber \\
 &  & +\sum_{m,d=1}^{M}\sum_{i=1}^{N_c}\left(b_{mi}^{\ast}b_{di}+b_{mi}^{\ast}B_{di}\sum_{k=1}^{N_b}g_{mk}^{\ast}\dot{g}_{dk}+B_{mi}^{\ast}b_{di}\sum_{k=1}^{N_b}\dot{g}_{mk}^{\ast}g_{dk}\right)\langle g_{m}|g_{d}\rangle\nonumber \\
 &  & +\sum_{m,d=1}^{M}\sum_{i=1}^{N_c}B_{mi}^{\ast}B_{di}\sum_{k,p=1}^{N_b}(\delta_{k,p}+g_{mk}^{\ast}g_{dp})\dot{g}_{mp}^{\ast}\dot{g}_{dp}\langle g_{m}|g_{d}\rangle.\label{eq:normdot}
\end{eqnarray}
\end{widetext}
The present variational approach can also be directly applied to the dissipative JC model by replacing $H_{\rm Rabi}$ with $H_{\rm JC}$ in Eqs. \eqref{eq:vecI1}-\eqref{eq:normdot}.

In Fig.~\ref{fig7}, we display the behavior of $\sigma^2(t)$ as a function of time $t$ for each variational result in the main text. According to our previous work~\cite{Yan_2020}, the variational dynamics agrees well with that of the hierarchical equations of motion (HEOM) approach~\cite{Tanimura2020,Moix2015} when $\sigma^2(t)<0.01$ and apparently deviates from the latter when $\sigma^2(t)\sim0.1$. Except for the case of $\alpha=0.2$ and the initial state being $|g,1_c\rangle$, one notes that $\sigma^2(t)$ is of order $10^{-3}$, which is much smaller than $0.01$. It is therefore reasonable to believe that the corresponding variational results are fairly accurate. For $\alpha=0.2$ and initial state being $|g,1_c\rangle$, although $\sigma^2(t)$ is a little larger than 0.01, we can also expect that the variational spectra slightly deviate from the exact ones since the deviation vector is still not too large. In addition, Fig.~\ref{fig7} indicates that the magnitude of $\sigma^2(t)$ depends on $\alpha$ and initial state.

\end{document}